\documentclass[conference]{IEEEtran}
\IEEEoverridecommandlockouts
\pdfoutput=1 
\usepackage{cite}
\usepackage{amsmath,amssymb,amsfonts}
\usepackage{algorithmic}
\usepackage{graphicx}

\usepackage{textcomp}
\usepackage{xcolor}
\usepackage{makecell}

\def\BibTeX{{\rm B\kern-.05em{\sc i\kern-.025em b}\kern-.08em
    T\kern-.1667em\lower.7ex\hbox{E}\kern-.125emX}}

\begin{document}
	
\bibliographystyle{IEEEtran}

\title{Channel Modeling for Heterogeneous Vehicular ISAC System with Shared Clusters}

\author{
\IEEEauthorblockN{Baiping Xiong\IEEEauthorrefmark{1}\IEEEauthorrefmark{2}, Zaichen Zhang\IEEEauthorrefmark{1}\IEEEauthorrefmark{2}, Yingmeng Ge\IEEEauthorrefmark{1}\IEEEauthorrefmark{2}, Haibo Wang\IEEEauthorrefmark{1}\IEEEauthorrefmark{2},  Hao Jiang\IEEEauthorrefmark{1}\IEEEauthorrefmark{3}, Liang Wu\IEEEauthorrefmark{1}\IEEEauthorrefmark{2}, and Ziyang Zhang\IEEEauthorrefmark{1}\IEEEauthorrefmark{2} }  

\IEEEauthorblockA{\IEEEauthorrefmark{1}National Mobile Communications Research Laboratory, Southeast University, Nanjing 210096, China}  
\IEEEauthorblockA{\IEEEauthorrefmark{2}Purple Mountain Laboratories, Nanjing 211111, China}  
\IEEEauthorblockA{\IEEEauthorrefmark{3}College of Artificial Intelligence, Nanjing University of Information Science and Technology, Nanjing 210044, China}  
\IEEEauthorblockA{Corresponding Author: Zaichen Zhang}  
\IEEEauthorblockA{Emails: \{xiongbp, zczhang, ymge, haibowang\}@seu.edu.cn, jianghao@nuist.edu.cn, \{wuliang, ziyangzhang\}@seu.edu.cn.}  
}

\maketitle

\begin{abstract}
In this paper, we consider the channel modeling of a heterogeneous vehicular integrated sensing and communication (ISAC) system, where a dual-functional multi-antenna base station (BS) intends to communicate with a multi-antenna vehicular receiver (MR) and sense the surrounding environments simultaneously. The time-varying complex channel impulse responses (CIRs) of the sensing and communication channels are derived, respectively, in which the sensing and communication channels are correlated with shared clusters. The proposed models show great generality for the capability in covering both monostatic and bistatic sensing scenarios, and as well for considering both static clusters/targets and mobile clusters/targets. Important channel statistical characteristics, including time-varying spatial cross-correlation function (CCF) and temporal auto-correlation function (ACF), are derived and analyzed. Numerically results are provided to show the propagation characteristics of the proposed ISAC channel model. Finally, the proposed model is validated via the agreement between theoretical and simulated as well as measurement results. 
\end{abstract}


\section{Introduction}

With the ongoing commercial deployment of fifth generation (5G) wireless network all over the world, the field results have shown the shortcomings of 5G in meeting with the increasing requirements of the future, and therefore researchers are devoting their attention to the next generation wireless network \cite{Walid6G}. Different from the previous generations of the wireless networks with one exclusive purpose of wireless communications, the future sixth generation (6G) wireless network is a heterogeneous network with an integration of various functions such as communications, sensing, computing, and controlling \cite{CYHISAC}. Specifically, the heterogeneous vehicular system provides low-latency high-throughput sensing and cooperation services, inspiring the research on the integrated sensing and communications (ISAC) technology from both academia and industry \cite{HeterogeneousV2X}.

The ISAC is a design methodology including associate enabling technologies that integrates wireless sensing and communication capabilities into one system to realize efficient utilization of hardware as well as radio resources and to achieve mutual benefits \cite{CYHISAC}. Specifically, the exploration of higher frequency bands as well as the ubiquitous devices enables the wireless system to better understand the surrounding environments through radio propagation. The precise sensing results of the targets as well as the physical environments, on the other hand, can contribute to efficient communication algorithms design. In \cite{ISACCAVmmWave}, the authors considered the dynamic frame structure design of a multi-vehicle ISAC system for the purpose of achieving low-latency and high-throughput sensing data sharing among vehicles, where results from the testbed validate the feasibility of the proposed design. The authors in \cite{machfilter} investigated the beamforming design of a ISAC vehicular system, where the results show that the communication beam tracking overheads could be significantly reduced with the assistance of sensing capability of ISAC.

For the development of ISAC, one of the indispensable aspects lies in the investigation of underlying propagation characteristics and development of appropriate channel models \cite{HybridRIS}. Thus far, only a few studies have involved in the ISAC channel research. The authors in \cite{DetermTHz} introduced the deterministic channel modeling approaches, including integral equation, physical optics, and geometrical optics, for sub-terahertz (THz) imaging scenarios, where the results show that the geometrical optics based modeling solution is more compatible with the measurements especially in three-dimensional (3D) imaging applications. In \cite{RSH_ISAC}, the authors developed a 3GPP-extended channel model for ISAC scenario, where both communication and sensing channels are generated based on the 3GPP structure with random parameters. The authors in \cite{YangrrISAC} proposed a 3D non-stationary ISAC channel model for monostatic sensing scenario, where a forward and backward scattering structure is considered for the communication channel. It is worth mentioning that the communication and sensing channels share the same space and frequency resources, indicating the existence of shared clusters/targets in ISAC channel. Different from communication channels with stochastically distributed clusters, the performance analysis of sensing channels heavily related to the deterministic locations of transceivers and clusters \cite{DetermTHz}. Also, the status of clusters/targets in ISAC channel show different impacts on the propagation properties.

To address the aforementioned issues, this paper develops a general 3D non-stationary multi-antenna heterogeneous vehicular ISAC channel model applicable for both monostatic and bistatic sensing scenarios, where the channel responses of communication and sensing channels are separately derived. To characterize the unique properties of the heterogeneous vehicular ISAC system, we adopt the geometrical optics based deterministic model for characterizing sensing channel and the geometrical based stochastic model for characterizing the communication channel, respectively. The communication and sensing channels are partially correlated by considering a shared cluster structure, where both static clusters/targets and mobile clusters/targets are considered. The well agreement between theoretical and simulated results as well as measurements validate the proposed model.

\emph{Notation:} In this paper, non-boldface, boldface lowercase, and boldface uppercase letters denote scalar, vector, and matrix, respectively; $\vert \cdot \vert$, $(\cdot)^{\ast}$, and $\langle\cdot , \cdot\rangle$ stand for absolute value, complex conjugate, and vector dot product, respectively; $\mathbb{E}\{\cdot\}$ and $\cap$ take the expectation and intersection set, respectively.

\begin{figure}[!t]  %
\centerline{\includegraphics[width=3.2in,height=1.8in]{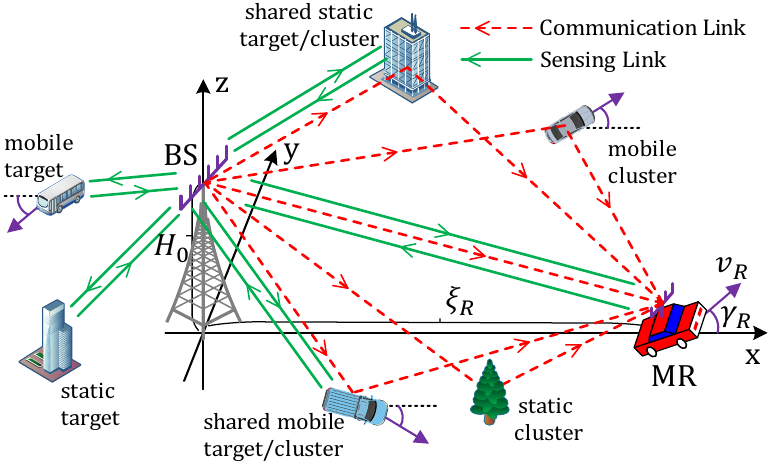}}
\caption{A physical illustration of the proposed heterogeneous vehicular ISAC system with shared targets/clusters, taking monostatic sensing as an example.}
\end{figure}

\section{System Model}

As illustrated in Fig. 1, we consider a heterogeneous vehicular integrated sensing and communication (ISAC) system consisting of a static dual-functional base station (BS) with height $H_0$ as well as a ground mobile receiver (MR) with velocity vector $\textbf{v}_R = v_R [\cos\gamma_R, \sin\gamma_R, 0]^{\emph{T}}$, where $v_R$ is the moving speed and $\gamma_R$ is the moving direction, respectively. The dual-functional BS intends to communicate with the MR and at the same time sensing the surrounding environment. Different from the conventional communication system \cite{MIMOComm} and/or radar sensing system \cite{MIMORadar}, where the signals are emitted by the BS with only one purpose in mind, either communicating with the receiver or sensing the target. In ISAC system, the signals transmitted by the BS are reused, taking the responsibilities for not only carrying the information bits to the receiver but also sensing the surrounding environment simultaneously. Therefore, the ISAC channel is consists of two parts, they are, the communication channel between the dual-functional BS and the communication receiver as well as the sensing channel between the dual-functional BS and the sensing targets, respectively. In this paper, the definition of the global coordinate system and that of the angular parameters in the coordinate system follow the procedure in \cite{X_ReLoS}, where the projection of the center of BS transmit antenna array is defined as the origin and the line connecting the origin and the center of communication receiver's antenna array is defined as the $x$-axis, respectively. For ease of clarification, all antenna arrays are assumed to be uniform linear arrays (ULAs) with omnidirectional radiating pattern. The definitions of the key model parameters, including parameters for sensing channel and communication channel, are summarized in Table I.

\begin{table}
\scriptsize     
\centering
\caption{Summary of Key Parameters Definitions}
\vspace*{-0.10cm}
\begin{tabular}{|c|c|}
\hline
$\xi^{T}_{\text{MR}}(t)$, $\xi^{R^e}_{\text{MR}}(t)$  &  distances between BS/echo receiver and MR   \\
\hline
$\xi^{T}_{s^{\ell_1}}$, $\xi^{R^e}_{s^{\ell_1}}$  &  distances between BS/echo receiver and $s_{\ell_1}$   \\
\hline
$\xi^{T}_{s^{\ell_2}}(t)$, $\xi^{R^e}_{s^{\ell_2}}(t)$  &  distances between BS/echo receiver and $s_{\ell_2}$   \\
\hline   
$\alpha^{i_2}_{T}(t)$, $\beta^{i_2}_{T}(t)$  &  AAoD/EAoD from BS to target $i_2 \in \big\{\text{MR}$, $s_{\ell_1}$, $s_{\ell_2}\big\}$   \\
\hline   
$\alpha^{i_2}_{R^e}(t)$, $\beta^{i_2}_{R^e}(t)$  &  AAoA/EAoA from target $i_2 \in \big\{\text{MR}$, $s_{\ell_1}$, $s_{\ell_2}\big\}$ to BS    \\
\hline 
$\xi^{T}_{s^{\ell_3}}$, $\xi^{R}_{s^{\ell_3}}(t)$  &  distances between BS/MR and $s_{\ell_3}$   \\
\hline
$\xi^{T}_{\ell_3, i_3}$, $\xi^{R}_{\ell_3, i_3}(t)$  &  distances between BS/MR and $s_{\ell_3}$ via $i_3$-th ray   \\
\hline
$\xi^{T}_{s^{\ell_4}}(t)$, $\xi^{R}_{s^{\ell_4}}(t)$  &  distances between BS/MR and $s_{\ell_4}$   \\
\hline
$\xi^{T}_{\ell_4, i_4}(t)$, $\xi^{R}_{\ell_4, i_4}(t)$  &  distances between BS/MR and $s_{\ell_4}$ via $i_4$-th ray   \\
\hline   
$\alpha^{\ell_3, i_3}_{T}$, $\beta^{\ell_3, i_3}_{T}$  &  AAoD/EAoD from BS to $s_{\ell_3}$ via $i_3$-th ray   \\
\hline   
$\alpha^{\ell_3, i_3}_{R}(t)$, $\beta^{\ell_3, i_3}_{R}(t)$  &  AAoA/EAoA from $s_{\ell_3}$ to MR via $i_3$-th ray    \\
\hline
$\alpha^{\ell_4, i_4}_{T}(t)$, $\beta^{\ell_4, i_4}_{T}(t)$  &  AAoD/EAoD from BS to $s_{\ell_4}$ via $i_4$-th ray      \\
\hline
$\alpha^{\ell_4, i_4}_{R}(t)$, $\beta^{\ell_4, i_4}_{R}(t)$  &  AAoA/EAoA from $s_{\ell_4}$ to MR via $i_4$-th ray     \\
\hline
\end{tabular}
\vspace*{-0.10cm}
\end{table}

In the ISAC system, the dual-functional BS emits the dual-functional signals for communication and sensing purpose. Some of the transmitted dual-functional signals will travel through the communication channel to the communication receiver, others will impinge on the sensing targets, getting reflected by the targets, and then travel to the sensing receiver via the sensing channel. Based on the physical location of the sensing antenna array, the sensing sub-system in ISAC system is generally categorized as monostatic sensing, also known as active sensing, and bistatic sensing, also known as passive sensing, respectively \cite{RCS}. In monostatic sensing, the antenna array is reused for both transmitting sensing signals and receiving the echo sensing signals, which is enabled by the development of full duplex antenna technology. In bistatic sensing, on the other hand, the sensing signal transmitting antenna array and the echo receiving antenna array are two different antenna arrays with physically separated locations, mounted either on the same platform or different platforms. To provide a general description, we assume the center of the sensing signal transmitting antenna array is mounted at position ($0, 0, H_0$) with antenna number denoted by $M_T$ and antenna element spacing denoted by $\delta_T$, respectively. The orientation angle of the sensing signal transmitting ULA is defined as $\psi_T$ in the azimuth direction and $\phi_T$ in the elevation direction, respectively. At the echo receiving side, the sensing signal receiving ULA is located at point ($x_{R^e}, y_{R^e}, z_{R^e}$) with antenna number denoted by $M_{R^e}$ and antenna element spacing denoted by $\delta_{R^e}$, respectively. Similarly, we use $\psi_{R^e}$ and $\psi_{R^e}$ to represent the azimuth orientation angle and elevation orientation angle of the echo receiving ULA, respectively. It can be seen that the proposed model can be adapted to characterize the monostatic sensing ISAC system when $x_{R^e} = 0$, $y_{R^e} = 0$, $z_{R^e} = H_0$, $M_{R^e} = M_T$, $\delta_{R^e} = \delta_T$, $\psi_{R^e} = \psi_T$, and $\phi_{R^e} = \phi_T$, respectively, otherwise, it characterizes a bistatic sensing ISAC system.

For the communication sub-system in the ISAC system, it shares the same transmitting ULA as the sensing sub-system, and therefore we use the same symbol notations for the transmit antenna parameters definition in the communication channel. As the communication receiver of the communication sub-system in the ISAC system is concerned, the communication signal receiving ULA and sensing signal receiving ULA are obviously not the same antenna array. In particular, the communication signal receiving ULA has an antenna number denoted by $M_R$ and an antenna element spacing denoted by $\delta_R$, respectively, with its orientation angle being denoted by $\psi_R$ and $\phi_R$ in the azimuth and elevation directions, respectively. Moreover, the location of the center of the communication signal receiving ULA is at point ($\xi_R, 0, 0$) at the initial instant, which yields to be ($\xi_R + v_R t \cos\gamma_R, v_R t \sin\gamma_R, 0$) after a moving time interval of $t$.

\begin{table}
\footnotesize  
\centering
\caption{RCS Values of Typical Targets}
\begin{tabular}{|c|c|c|}
\hline
\textbf{Target Type}  &\hspace*{-0.375cm} & \textbf{RCS} [$\text{m}^2$]   \\
\hline
{\makecell[c]{ Automobile }}  &\hspace*{-0.375cm} &  100     \\
\hline
{\makecell[c]{ Pickup truck }}   &\hspace*{-0.375cm}  &  200   \\
\hline
{\makecell[c]{ Adult }}   &\hspace*{-0.375cm}  &  1   \\
\hline
{\makecell[c]{ Bird }}   &\hspace*{-0.375cm}  &  0.01   \\
\hline
{\makecell[c]{ Insect }}   &\hspace*{-0.375cm}  &  $10^{-5}$   \\
\hline
{\makecell[c]{ Missile }}   &\hspace*{-0.375cm}  &  0.5   \\
\hline
{\makecell[c]{ Jumbo jet airliner }}   &\hspace*{-0.375cm}  &  100   \\
\hline
{\makecell[c]{ Large bomber }}   &\hspace*{-0.375cm}  &  40   \\
\hline
{\makecell[c]{ Small fighter aircraft }}   &\hspace*{-0.375cm}  &  2   \\
\hline
\end{tabular}
\end{table}

\subsection{Sensing Channel Modeling}

In the sensing channel, the sensing targets span a wide range from the surrounding buildings, trees, and running vehicles, to the mobile communication receiver. In this case, we divide the targets into three categories, they are, the terminal target (e.g., communication receiver MR), the static targets (e.g., static buildings and trees, etc.), and the mobile targets (e.g., running vehicles and pedestrians, etc.), respectively. More specifically, we assume there are $L_1$ static targets forming a target set of $C_{L_1}$ with the $\ell_1$-th ($\ell_1 = 1, 2, ..., L_1$) target being denoted by $s_{\ell_1}$, and assume there are $L_2$ mobile targets on the ground surface forming a target set of $C_{L_2}$. For the $\ell_2$-th ($\ell_2 = 1, 2, ..., L_2$) mobile target in target set $C_{L_2}$, that is, $s_{\ell_2}$, its velocity vector is represented by $\textbf{v}_{\ell_2} = v_{\ell_2} [\cos\gamma_{\ell_2}, \sin\gamma_{\ell_2}, 0]^{\emph{T}}$, in which $v_{\ell_2}$ and $\gamma_{\ell_2}$ denote the moving speed and direction of the target $s_{\ell_2}$, respectively. Let $\textbf{s}(t) \in \mathbb{C}^{M_T \times 1}$ denote the dual-functional transmitted signal vector and $\textbf{y}^{e}(t) \in \mathbb{C}^{M_{R^e} \times 1}$ denote the received echo signal vector, respectively. Then, the received signal model for the sensing sub-system in the ISAC system can be expressed as \cite{machfilter}, \cite{RCS}
\begin{equation}
\setlength\abovedisplayskip{3pt}
\setlength\belowdisplayskip{2pt}
\begin{aligned}
\textbf{y}^{e} (t)  =   \textbf{H}^{e} (t, f) \textbf{s}(t) +  \textbf{n}^{e} (t) ,
\end{aligned}
\l
\end{equation}
where $\textbf{n}^{e}(t) \in \mathbb{C}^{M_{R^e} \times 1}$ is the noise component, $\textbf{H}^{e}(t, f) = \int \textbf{H}^{e}(t, \tau) e^{- j 2\pi f \tau} d \tau$ represents the frequency domain sensing channel matrix with $\textbf{H}^{e}(t, \tau) = \big[ h^{e}_{p q^e}(t, \tau) \big]_{M_{R^e} \times M_T}$ denoting its time domain counterpart. In particular, $h^{e}_{p q^e}(t, \tau)$ is the complex channel impulse response (CIR) including path loss between the $p$-th ($p = 1, 2, ..., M_T$) BS transmit antenna and the $q^e$-th ($q^e = 1, 2, ..., M_{R^e}$) echo receiving antenna in the sensing sub-system, which is composed of three components
\begin{eqnarray}
h^{e}_{pq^e}(t, \tau)   =   h^{e, \text{MR}}_{pq^e}(t, \tau)   +   h^{e, \text{static}}_{pq^e}(t, \tau)   +   h^{e, \text{mobile}}_{pq^e}(t, \tau) , 
\end{eqnarray}
where $h^{e, \text{MR}}_{pq^e}(t, \tau)$, $h^{e, \text{static}}_{pq^e}(t, \tau)$, and $h^{e, \text{mobile}}_{pq^e}(t, \tau)$ are the CIRs of the links corresponding to terminal target MR, static targets in set $C_{L_1}$, and mobile targets in set $C_{L_2}$, respectively. Here we have to mention that the CIR of the link for static target in set $C_{L_1}$ is time-invariant, thus we can simplify $h^{e, \text{static}}_{pq^e}(t, \tau)$ into $h^{e, \text{static}}_{pq^e}(\tau)$. Furthermore, their expressions are presented~as
\allowdisplaybreaks[4]
\begin{eqnarray}
h^{e, \text{MR}}_{pq^e}(t, \tau)   \hspace*{-0.225cm}&=&\hspace*{-0.225cm}   \sqrt{ \Omega^e_{\text{MR}}(t) }  e^{-j \frac{2\pi}{\lambda} ( \xi^T_{\text{MR}}(t) + \xi^{R^e}_{\text{MR}}(t) ) }  \nonumber \\ [0.15cm]
&&\hspace*{-0.225cm}\times   e^{ j \frac{2\pi}{\lambda} \langle \emph{\textbf{e}}^{\text{MR}}_{T}(t), \;  \textbf{d}^{T}_{p}  \rangle }     e^{ j \frac{2\pi}{\lambda} \langle \emph{\textbf{e}}^{\text{MR}}_{R^e}(t), \;  \textbf{d}^{R^e}_{q^e}  \rangle }  \nonumber \\ [0.15cm] 
&&\hspace*{-0.225cm}\times  e^{ j \frac{2\pi}{\lambda} \langle  - \textbf{v}_R t, \;  \emph{\textbf{e}}^{\text{MR}}_{T}(t)  \rangle }  e^{ j \frac{2\pi}{\lambda} \langle  - \textbf{v}_R t, \;  \emph{\textbf{e}}^{\text{MR}}_{R^e}(t)  \rangle }   \nonumber \\ [0.15cm]
&&\hspace*{-0.225cm}\times   \delta\big(\tau -   \big(\xi^T_{\text{MR}}(t) + \xi^{R^e}_{\text{MR}}(t)\big)/c \big)  ,
\end{eqnarray}
\begin{eqnarray}
h^{e, \text{static}}_{pq^e}( \tau)   \hspace*{-0.225cm}&=&\hspace*{-0.225cm}  \sum^{L_1}_{\ell_1 = 1} \sqrt{ \Omega^e_{s_{\ell_1}} }  e^{-j \frac{2\pi}{\lambda} ( \xi^T_{s_{\ell_1}} + \xi^{R^e}_{s_{\ell_1}} ) }  \nonumber \\ [0.15cm]
&&\hspace*{-0.225cm}\times   e^{ j \frac{2\pi}{\lambda} \langle \emph{\textbf{e}}^{s_{\ell_1}}_{T}, \;  \textbf{d}^{T}_{p}  \rangle }     e^{ j \frac{2\pi}{\lambda} \langle \emph{\textbf{e}}^{s_{\ell_1}}_{R^e}, \;  \textbf{d}^{R^e}_{q^e}  \rangle }  \nonumber \\ [0.15cm] 
&&\hspace*{-0.225cm}\times   \delta\big(\tau -   (\xi^T_{s_{\ell_1}} + \xi^{R^e}_{s_{\ell_1}})/c \big)     ,
\end{eqnarray}
\begin{eqnarray}
h^{e, \text{mobile}}_{pq^e}(t, \tau)   \hspace*{-0.225cm}&=&\hspace*{-0.225cm}  \sum^{L_2}_{\ell_2 = 1} \sqrt{ \Omega^e_{s_{\ell_2}}(t) }  e^{-j \frac{2\pi}{\lambda} ( \xi^T_{s_{\ell_2}}(t) + \xi^{R^e}_{s_{\ell_2}}(t) ) }   \nonumber \\ [0.15cm]
&&\hspace*{-0.225cm}\times   e^{ j \frac{2\pi}{\lambda} \langle \emph{\textbf{e}}^{s_{\ell_2}}_{T}(t), \;  \textbf{d}^{T}_{p}  \rangle }     e^{ j \frac{2\pi}{\lambda} \langle \emph{\textbf{e}}^{s_{\ell_2}}_{R^e}(t), \;  \textbf{d}^{R^e}_{q^e}  \rangle }  \nonumber \\ [0.15cm] 
&&\hspace*{-0.225cm}\times  e^{ j \frac{2\pi}{\lambda} \langle  - \textbf{v}_{\ell_2} t, \;  \emph{\textbf{e}}^{s_{\ell_2}}_{T}(t)  \rangle }  e^{ j \frac{2\pi}{\lambda} \langle  - \textbf{v}_{\ell_2} t, \;  \emph{\textbf{e}}^{s_{\ell_2}}_{R^e}(t)  \rangle }   \nonumber \\ [0.15cm]
&&\hspace*{-0.225cm}\times   \delta\big(\tau -   \big(\xi^T_{s_{\ell_2}}(t) + \xi^{R^e}_{s_{\ell_2}}(t)\big)/c \big)     ,
\end{eqnarray}
where $c = 3.0 \times 10^8$ m/s, $\Omega^e_{\text{MR}}(t) = \frac{\lambda^2 \sigma_{\text{MR}} }{(4\pi)^3 (\xi^T_{\text{MR}}(t) \xi^{R^e}_{\text{MR}}(t))^2 }$, $\Omega^e_{s_{\ell_1}} = \frac{\lambda^2 \sigma_{s_{\ell_1}} }{(4\pi)^3 (\xi^T_{s_{\ell_1}} \xi^{R^e}_{s_{\ell_1}})^2}$, and $\Omega^e_{s_{\ell_2}}(t) = \frac{\lambda^2 \sigma_{s_{\ell_2}} }{(4\pi)^3 (\xi^T_{s_{\ell_2}}(t) \xi^{R^e}_{s_{\ell_2}}(t))^2}$ denote the path power gains including the path loss of the sensing links for target MR, static target $s_{\ell_1}$, and mobile target $s_{\ell_2}$, respectively. The $\sigma_{\text{MR}}$, $\sigma_{s_{\ell_1}}$, and $\sigma_{s_{\ell_2}}$ are the radar cross section (RCS) of the corresponding targets, whose typical values are given in Table II \cite{RCS}. The $\textbf{d}^T_p$ and $\textbf{d}^{R^e}_{q^e}$ represent the distance vectors from the centers of the signal transmitting and echo receiving ULAs to the $p$-th transmit and $q^e$-th echo receive antennas, respectively, and they are expressed as
\begin{eqnarray}
\hspace*{-0.4cm}  \textbf{d}^{T/R^e}_{i_1}  \hspace*{-0.245cm}&=&\hspace*{-0.245cm}  \frac{M_{T/R^e} - 2 i_1 + 1}{2} \delta_{T/R^e} \hspace*{-0.125cm} 
\begin{bmatrix}
\cos\phi_{T/R^e} \cos\psi_{T/R^e}    \\ 
\cos\phi_{T/R^e} \sin\psi_{T/R^e}     \\ 
\sin\phi_{T/R^e}   
\end{bmatrix}  \hspace*{-0.125cm} ,
\end{eqnarray}
where $i_1 = p$ for transmit antenna, e.g., $\textbf{d}^{T}_{p}$, and $i_1 = q^e$ for echo receive antenna, e.g., $\textbf{d}^{R^e}_{q^e}$, respectively. In addition, $\big\{\emph{\textbf{e}}^{\text{MR}}_{T/R^e}(t), \emph{\textbf{e}}^{s_{\ell_1}}_{T/R^e}, \emph{\textbf{e}}^{s_{\ell_2}}_{T/R^e}(t)\big\}$ denote the unit directional vectors from transmitting/echo receiving ULA to target MR, static target $s_{\ell_1}$, and mobile target $s_{\ell_2}$, respectively, which contain the location information of the targets and can be expressed~as
\begin{eqnarray}
\emph{\textbf{e}}^{i_2}_{T/R^e}(t)  \hspace*{-0.225cm}&=&\hspace*{-0.225cm} 
\begin{bmatrix}
\hspace*{0.05cm}  \cos\beta^{i_2}_{T/R^e}(t) \cos\alpha^{i_2}_{T/R^e}(t)  \hspace*{0.05cm}  \\[0.05cm]
\hspace*{0.05cm}  \cos\beta^{i_2}_{T/R^e}(t) \sin\alpha^{i_2}_{T/R^e}(t)   \hspace*{0.05cm}   \\[0.05cm]
\hspace*{0.05cm}  \sin\beta^{i_2}_{T/R^e}(t)   \hspace*{0.05cm}
\end{bmatrix} ,
\end{eqnarray}
where $i_2 \in \{\text{MR}, s_{\ell_1}, s_{\ell_2}\}$, and for static target $s_{\ell_1}$ we have $\alpha^{s_{\ell_1}}_{T/R^e}(t) = \alpha^{s_{\ell_1}}_{T/R^e}$ and $\beta^{s_{\ell_1}}_{T/R^e}(t) = \beta^{s_{\ell_1}}_{T/R^e}$, respectively.

It is worth mentioning that the CIRs in (2)-(5) is a general description of the channel response of the sensing sub-system in ISAC system, which can be adapted for characterizing both monostatic and bistatic sensing scenarios. Specifically, in monostatic sensing scenario, the echo links from sensing targets to echo receiving ULA share the same distance and angle parameters as the transmission links from transmitting ULA to the sensing targets, indicting that $\xi_{R^e, \text{MR}}(t) = \xi_{T, \text{MR}}(t)$, $\xi_{R^e, s_{\ell_1}} = \xi_{T, s_{\ell_1}}$, $\xi_{R^e, s_{\ell_2}}(t) = \xi_{T, s_{\ell_2}}(t)$, $\alpha^{\text{MR}}_{T}(t) = \alpha^{\text{MR}}_{R^e}(t)$, $\beta^{\text{MR}}_{T}(t) = \beta^{\text{MR}}_{R^e}(t)$, $\alpha^{s_{\ell_1}}_{T} = \alpha^{s_{\ell_1}}_{R^e}$, $\beta^{s_{\ell_1}}_{T} = \beta^{s_{\ell_1}}_{R^e}$, $\alpha^{s_{\ell_2}}_{T}(t) = \alpha^{s_{\ell_2}}_{R^e}(t)$, and $\beta^{s_{\ell_2}}_{T}(t) = \beta^{s_{\ell_2}}_{R^e}(t)$, respectively. Furthermore, by exploiting the matched-filtering or other advanced estimation technologies \cite{machfilter}, it is reasonable to assume that the distance and angle parameters of the targets can be well sensed.

\subsection{Communication Channel Modeling}

In the communication channel, the transmitted signals from the BS will travel through LoS path, NLoS paths with static clusters, and NLoS paths with mobile clusters to the communication receiver MR. We assume there are $L_3$ static clusters forming a cluster set of $C_{L_3}$, in which the $\ell_3$-th ($\ell_3 = 1, 2, ..., L_3$) cluster is denoted by $s_{\ell_3}$, and assume there are $L_4$ mobile clusters forming a cluster set of $C_{L_4}$ with the $\ell_4$-th ($\ell_4 = 1, 2, ..., L_4$) cluster being denoted by $s_{\ell_4}$, respectively. Each cluster contributes a multipath propagation link with $I$ rays. The velocity vector of the mobile cluster $s_{\ell_4}$ is $\textbf{v}_{\ell_4} = v_{\ell_4} [\cos\gamma_{\ell_4}, \sin\gamma_{\ell_4}, 0]^{\emph{T}}$ with $v_{\ell_4}$ and $\gamma_{\ell_4}$ denoting the moving speed and direction, respectively. We denote $\textbf{y}^c(t) \in \mathbb{C}^{M_{R} \times 1}$ as the MR received signal vector, $\textbf{n}^c(t) \in \mathbb{C}^{M_{R} \times 1}$ as the noise vector, and $\textbf{H}^{c}(t, f) = \int \textbf{H}^{c}(t, \tau) e^{- j 2\pi f \tau} d \tau$ as the frequency domain communication channel matrix, respectively. The received signal model for the communication sub-system in the ISAC system yields to be \cite{MIMOComm}, \cite{X_ReLoS}
\begin{eqnarray}
\textbf{y}^{c} (t) \hspace*{-0.225cm}&=&\hspace*{-0.225cm}   \int \textbf{H}^{c}(t, \tau) e^{- j 2\pi f \tau} d \tau  \;  \textbf{s}(t) +  \textbf{n}^{c} (t)  ,
\end{eqnarray}
where $\textbf{H}^{c}(t, \tau) = \big[ h^c_{p q}(t, \tau) \big]_{M_{R} \times M_T}$ is the time domain communication channel matrix and $h^c_{p q}(t, \tau)$ denotes the CIR between the $p$-th BS transmit antenna and $q$-th ($q = 1, 2, ..., M_R$) MR receive antenna in the communication sub-system, i.e., 
\allowdisplaybreaks[4]
\begin{eqnarray}
h^{c}_{pq}(t, \tau)  \hspace*{-0.225cm}&=&\hspace*{-0.225cm}  \sqrt{\Omega^c_{\text{LoS}}(t)}  h^{c, \text{LoS}}_{pq}(t)  \delta\big(\tau - \tau_{\text{LoS}}(t) \big)  \nonumber \\ [0.15cm]
&&\hspace*{-0.225cm}+  \sum^{L_3}_{\ell_3 = 1} \sqrt{\Omega^c_{s_{\ell_3}}(t)}  h^{c, s_{\ell_3}}_{pq}(t)  \delta\big(\tau - \tau_{\ell_3}(t) \big)    \nonumber \\ [0.15cm]
&&\hspace*{-0.225cm}+  \sum^{L_4}_{\ell_4 = 1} \sqrt{\Omega^c_{s_{\ell_4}}(t)}  h^{c, s_{\ell_4}}_{pq}(t)  \delta\big(\tau - \tau_{\ell_4}(t) \big)  ,
\end{eqnarray}
where $\tau_{\text{LoS}}(t) = \xi^{T}_{\text{MR}}(t)/c$, $\tau_{\ell_3}(t) = \big(\xi^T_{\ell_3} + \xi^R_{\ell_3}(t) \big)/c$, and $\tau_{\ell_4}(t) = \big(\xi^T_{\ell_4}(t) + \xi^R_{\ell_4}(t)\big)/c$ denote the propagation delays of the LoS path, NLoS path via static cluster $s_{\ell_3}$, and NLoS path via mobile cluster $s_{\ell_4}$, respectively. The $\Omega^c_{\text{LoS}}(t) = \frac{\lambda^2}{(4\pi)^2 {\xi^T_{\text{MR}}}^2(t)}$, $\Omega^c_{s_{\ell_3}}(t) = \frac{\lambda^2 P_{\ell_3}(t)}{(4\pi)^2 ( \xi^T_{\ell_3} + \xi^R_{\ell_3}(t) )^2}$, and $\Omega^c_{s_{\ell_4}}(t) = \frac{\lambda^2 P_{\ell_4}(t)}{(4\pi)^2 ( \xi^T_{\ell_4}(t) + \xi^R_{\ell_4}(t) )^2}$ represent the path power gains including path loss of the LoS path, NLoS path via static cluster $s_{\ell_3}$, and NLoS path via mobile cluster $s_{\ell_4}$, respectively, in which $P_{\ell_3}(t)$ and $P_{\ell_4}(t)$ denote the powers of clusters $s_{\ell_3}$ and $s_{\ell_4}$, respectively \cite{2RISTCOM}, \cite{3GPP}. Furthermore, $h^{c, \text{LoS}}_{pq}(t)$, $h^{c, s_{\ell_3}}_{pq}(t)$, and $h^{c, s_{\ell_4}}_{pq}(t)$ are the CIRs between the ($p, q$)-th transmit-receive antenna pair of the LoS path, NLoS path via static cluster $s_{\ell_3}$, and NLoS path via mobile cluster $s_{\ell_4}$, respectively, and they can be expressed as
\begin{eqnarray}
h^{c, \text{LoS}}_{pq}(t)   \hspace*{-0.225cm}&=&\hspace*{-0.225cm}   e^{-j \frac{2\pi}{\lambda} \xi^T_{\text{MR}}(t) }  \times   e^{ j \frac{2\pi}{\lambda} \langle \emph{\textbf{e}}^{\text{MR}}_{T}(t), \;  \textbf{d}^{T}_{p}  \rangle }    \nonumber \\ [0.15cm] 
&&\hspace*{-0.225cm}\times  e^{ j \frac{2\pi}{\lambda} \langle  - \emph{\textbf{e}}^{\text{MR}}_{T}(t), \;  \textbf{d}^{R}_{q}  \rangle }   \times   e^{ j \frac{2\pi}{\lambda} \langle  \textbf{v}_R t, \;  - \emph{\textbf{e}}^{\text{MR}}_{T}(t)  \rangle }   ,
\end{eqnarray}
\begin{eqnarray}
h^{c, s_{\ell_3}}_{pq}(t)   \hspace*{-0.225cm}&=&\hspace*{-0.225cm}   \sqrt{ \frac{1}{I} } \sum^{I}_{i_3 = 1} e^{j \big( \varphi_{\ell_3, i_3} - \frac{2\pi}{\lambda} \big( \xi^T_{\ell_3, i_3} + \xi^R_{\ell_3, i_3}(t) \big) \big) }    \nonumber \\ [0.15cm]
&&\hspace*{-0.225cm}\times   e^{ j \frac{2\pi}{\lambda} \langle \emph{\textbf{e}}^{\ell_3, i_3}_{T}, \;  \textbf{d}^{T}_{p}  \rangle }   \times   e^{ j \frac{2\pi}{\lambda} \langle \emph{\textbf{e}}^{\ell_3, i_3}_{R}(t), \;  \textbf{d}^{R}_{q}  \rangle }   \nonumber \\ [0.15cm]
&&\hspace*{-0.225cm}\times  e^{ j \frac{2\pi}{\lambda} \langle  \textbf{v}_R t, \;  \emph{\textbf{e}}^{\ell_3, i_3}_{R}(t)  \rangle }  ,
\end{eqnarray}
\begin{eqnarray}
\hspace*{-0.5cm} h^{c, s_{\ell_4}}_{pq}(t)   \hspace*{-0.225cm}&=&\hspace*{-0.225cm}  \sqrt{ \frac{1}{I} }  \sum^{I}_{i_4 = 1}  e^{ j \big( \varphi_{\ell_4, i_4} -  \frac{2\pi}{\lambda} \big( \xi^T_{\ell_4, i_4}(t)  +  \xi^R_{\ell_4, i_4}(t) \big) \big) }   \nonumber \\ [0.15cm]
&&\hspace*{-0.315cm}\times  e^{ j \frac{2\pi}{\lambda} \langle \emph{\textbf{e}}^{\ell_4, i_4}_{T}(t), \;  \textbf{d}^{T}_{p}  \rangle }   \times  e^{ j \frac{2\pi}{\lambda} \langle  - \textbf{v}_{\ell_4} t, \;  \emph{\textbf{e}}^{\ell_4, i_4}_{T}(t)  \rangle }    \nonumber \\ [0.15cm] 
&&\hspace*{-0.315cm}\times  e^{ j \frac{2\pi}{\lambda} \langle \emph{\textbf{e}}^{\ell_4, i_4}_{R}(t), \;  \textbf{d}^{R}_{q}  \rangle }  \times  e^{ j \frac{2\pi}{\lambda} \langle  ( \textbf{v}_R - \textbf{v}_{\ell_4} ) t, \;  \emph{\textbf{e}}^{\ell_4, i_4}_{R}(t)  \rangle }  , 
\end{eqnarray}
where $\big\{\varphi_{\ell_3, i_3}\big\}^{\ell_3 = 1, ..., L_3}_{i_3 = 1, ..., I}$ and $\big\{\varphi_{\ell_4, i_4}\big\}^{\ell_4 = 1, ..., L_4}_{i_4 = 1, ..., I}$ are assumed to be independent and uniformly distributed random phases. $\textbf{d}^R_q = \frac{M_R - 2q + 1}{2}\delta_R [\cos\phi_R\cos\psi_R, \cos\phi_R\sin\psi_R, \sin\phi_R ]^{\emph{T}}$. Moreover, $\emph{\textbf{e}}^{\ell_3, i_3}_{T}$ and $\emph{\textbf{e}}^{\ell_3, i_3}_{R}(t)$ denote the unit directional vectors from BS transmitting ULA and MR receiving ULA to the static cluster $s_{\ell_3}$ via the $i_3$-th ($i_3 = 1, ..., I$) ray, respectively; whereas $\emph{\textbf{e}}^{\ell_4, i_4}_{T}(t)$ and $\emph{\textbf{e}}^{\ell_4, i_4}_{R}(t)$ are the uni directional vectors from BS transmitting ULA and MR receiving ULA to the mobile cluster $s_{\ell_4}$ via the $i_4$-th ($i_4 = 1, ..., I$) ray, respectively. Their expressions can be obtained from $\emph{\textbf{e}}^{\text{MR}}_{T}(t)$ in (7) by replacing $\big\{\alpha^{\text{MR}}_{T}(t), \beta^{\text{MR}}_{T}(t)\big\}$ with $\big\{\alpha^{\ell_3, i_3}_{T}, \beta^{\ell_3, i_3}_{T}\big\}$, $\big\{\alpha^{\ell_3, i_3}_{R}(t), \beta^{\ell_3, i_3}_{R}(t)\big\}$, $\big\{\alpha^{\ell_4, i_4}_{T}(t), \beta^{\ell_4, i_4}_{T}(t)\big\}$, and $\big\{\alpha^{\ell_4, i_4}_{R}(t), \beta^{\ell_4, i_4}_{R}(t)\big\}$, respectively, which are omitted here for brevity.

Since the communication receiver MR is sensed as a target in the sensing channel, the time-varying model parameters of the LoS path can be provided by the sensing channel. For the NLoS paths, including NLoS paths via static and mobile clusters, the time-varying model parameters are obtained according to the geometrical relationship among transmit ULA, clusters, and receive ULA, based on the initial location information of the clusters as well as the motion parameters of the MR and/or clusters, in which the initial location information of the clusters are generally generated randomly \cite{V2X}. In this paper, however, we consider the sensing channel and communication channel share some common clusters, and therefore the initial location information of the shared clusters could be provided by the sensing channel. For arbitrary shared cluster $s_{\ell_5}$, i.e., $s_{\ell_5} \in \big\{C_{L_1} \cap C_{L_3}\big\}$ for shared static clusters or $s_{\ell_5} \in \big\{C_{L_2} \cap C_{L_4}\big\}$ for shared mobile clusters, the communication channel and sensing channel share the same propagation from transmit ULA to the shared cluster $s_{\ell_5}$, indicating that the location information of $s_{\ell_5}$ could be sensed. Then, based on the sensed location information of the shared cluster $s_{\ell_5}$, the time-varying distance and angle parameters of the propagation link via the shared cluster $s_{\ell_5}$ could be updated following the same procedure in \cite{2RISTCOM}, \cite{V2X}, \cite{X_mmWaveUAVRIS}. The detailed expressions are omitted here due to space limitation.

\section{Channel Characteristics}

\subsection{Time-varying Spatial CCF}

The time-varying spatial cross-correlation function (CCF) characterize the correlation properties of the channel between two different links in the space domain, which is widely used as a metric to measure the spatial diversity of multi-antenna channels and is defined as \cite{X_mmWaveUAVRIS}
\begin{eqnarray}
\rho_{(p,q),(p',q')}(t, \Delta p, \Delta q)  =   \frac{\mathbb{E}\big[h^\ast_{pq}(t) h_{p'q'}(t) \big]}{\sqrt{\mathbb{E}\big[\vert h_{pq}(t) \vert^2 \big] \mathbb{E}\big[\vert h_{p'q'}(t) \vert^2 \big] }}   ,
\end{eqnarray}
where $\Delta p = \vert p' - p \vert \delta_T/\lambda$ and $\Delta q = \vert q' ({q^e}') - q ({q^e}) \vert \delta_{R (R^e)}/\lambda$ represent the normalized transmit antenna spacing and normalized receive antenna spacing, respectively. By substituting $h^e_{p q^e}(t, \tau)$ from (2)-(5) and $h^c_{pq}(t, \tau)$ from (9)-(12) into (13), we can obtain the normalized spatial CCF of the sensing channel $\rho^{e}_{(p,q^e),(p',{q^e}')}(t, \Delta p, \Delta q^e)$ and that of the communication channel $\rho^c_{(p,q),(p',q')}(t, \Delta p, \Delta q)$, respectively. Their expressions are omitted here due to the space limitation.

\subsection{Time-varying Temporal ACF}

The time-varying temporal auto-correlation function (ACF) measures the correlation ratio of the link at two different time instants, where a smaller value of temporal ACF means a faster changing of the channel and hence requiring a more frequently estimation of the channel. The temporal ACF is defined as \cite{X_mmWaveUAVRIS}
\begin{eqnarray}
\rho_{(p,q)}(t, \Delta t)  =   \frac{\mathbb{E}\big[h^\ast_{pq}(t) h_{pq}(t + \Delta t) \big]}{\sqrt{\mathbb{E}\big[\vert h_{pq}(t) \vert^2 \big] \mathbb{E}\big[\vert h_{pq}(t + \Delta t) \vert^2 \big] }}   ,
\end{eqnarray}
where $\Delta t$ is the time difference. The corresponding expression of the temporal ACF of the sensing channel and that of the communication channel can be obtained by substituting $h^e_{p q^e}(t, \tau)$ from (2)-(5) and $h^c_{pq}(t, \tau)$ from (9)-(12) into (14), respectively, whose expressions are omitted here for brevity.

\section{Results and Discussions}

In this section we numerically investigate the propagation characteristics of the proposed heterogeneous vehicular ISAC channel model. The ISAC system operates in the mmWave frequency band with carrier frequency of $f_c = 28$ GHz, the parameter setting for the dual-function BS follows $H_0 = 30$ m, $M_T = 4$, $\delta_T = \lambda/2$, $\psi_T = \pi/3$, and $\phi_T = \pi/4$, respectively. For the communication sub-system, we set $\xi_R = 150$ m, $M_R = 6$, $\delta_R = \lambda/2$, $\psi_R = \pi/4$, and $\phi_R = \pi/4$, respectively. Moreover, the communication receiver MR is in motion with speed $v_R = 5$ m/s and direction $\gamma_R = - \pi/6$, respectively. As the sensing sub-system is concerned, in monostatic sensing scenario the sensing echo receiver shares the same hardware as the dual-functional BS, thus having the same parameter setting. In bistatic sensing scenario, on the other hand, we set $x_{R^e} = 100$ m, $y_{R^e} = - 30$ m, and $z_{R^e} = 30$ m, respectively, and assume that $M_{R^e} = 4$, $\delta_{R^e} = \lambda/2$, $\psi_{R^e} = \pi/3$, and $\phi_{R^e} = \pi/4$, respectively.

\begin{figure}[!t]  %
\centerline{\includegraphics[width=0.39\textwidth]{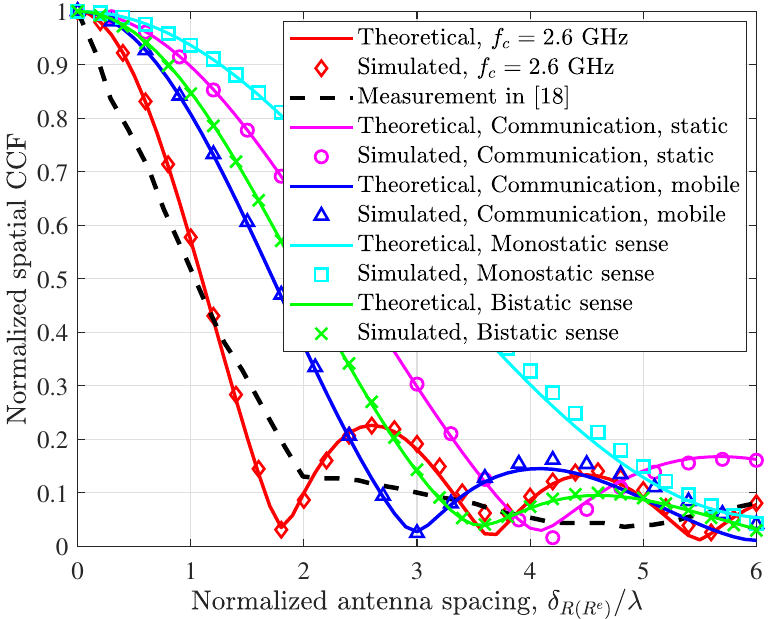}}
\caption{Spatial CCF of the proposed ISAC channel model when $t = 2$ s, where `Communication, static' means communication channel with static clusters.}
\end{figure}

By exploiting (13), Figure 2 compares the normalized spatial CCF of the proposed ISAC channel model in different components when $t = 2$ s, where the results show that the spatial correlations of both communication channel and sensing channel decrease gradually as the antenna spacing increases \cite{YangrrISAC}. The results indicate that the communication channel and sensing channel show different spatial correlation properties even when the shared cluster is considered, which could be interpreted by the fact that the propagations from the shared cluster to the receivers in communication and sensing channels are independent. Moreover, Fig. 2 shows that the communication channel with mobile clusters show faster decline of the spatial CCF as compared to that with static clusters, which is mainly because that more dynamics involved in the channel will cause faster vary of the channel and thus helps improve the channel spatial diversity. In addition, it is seen from Fig. 2 that the bistatic sensing channel show smaller spatial correlation than monostatic sensing channel, thus is more benefit for channel spatial diversity. It is also seen from Fig. 2 that the simulated results match with the theoretical ones well, which confirms the correctness of the derived spatial CCF of the proposed ISAC channel model; meanwhile, the well agreement between measurement results from \cite{CCFMeasure} and theoretical as well as simulated ones under $f_c = 2.6$ GHz highlights the accuracy of the proposed channel model.

\begin{figure}[!t]    %
\centerline{\includegraphics[width=0.39\textwidth]{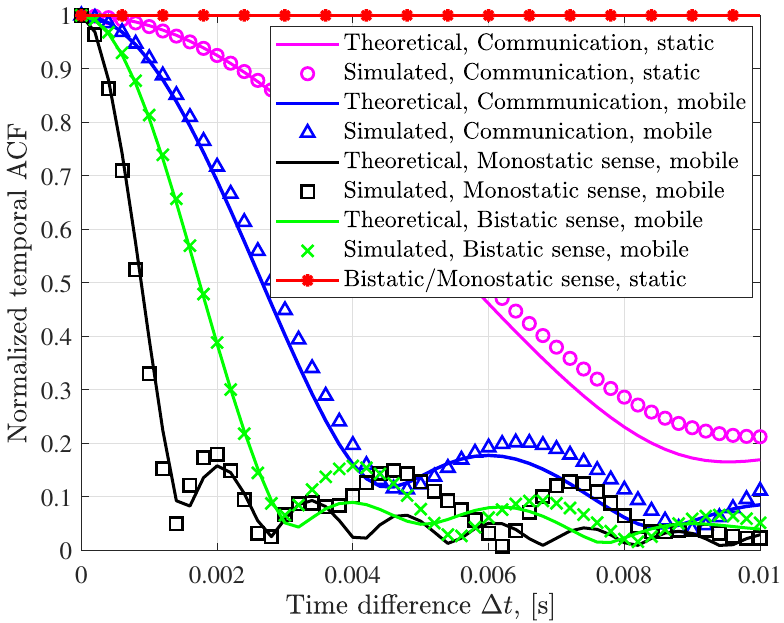}}
\caption{Temporal ACF of the proposed ISAC channel model when $t = 5$ s, where `Communication, static' means communication channel with static clusters.}
\end{figure}

In Figure 3, we study the temporal correlation properties of the proposed ISAC channel model for different propagation links at $t = 5$ s based on (14). The results show that the sensing channel with mobile targets/clusters show faster decline of the temporal ACF than communication channel, which means the sensing channel is more sensitive to the targets/clusters motion and changes more rapidly, arising for more frequent estimation of the channel. Also, it reveals that the monostatic sensing channel show smaller temporal correlation than bisttaic sensing channel, requiring more transmission resources for channel estimation, which should be well balanced in resource-limited ISAC system design. As for the communication channel, it is seen from Fig. 3 that the channel with mobile clusters has smaller and faster decreasing of temporal ACF as it involves more dynamics, which is in agreement with our previous results in \cite{V2X}. Finally, the excellent agreement between simulated and theoretical curves verify the effectiveness of the proposed ISAC channel model.

\section{Conclusion}

In this paper, we have considered the channel modeling of a multi-antenna heterogeneous vehicular ISAC system, where a dual-functional BS communicates with the MR and senses the surrounding environments simultaneously. By adopting the cluster-based structure, the channel responses of the sensing channel and communication channel considering the presence of both static as well as mobile targets/clusters are respectively derived, which are partially correlated with each other based on the shared targets/clusters. The proposed ISAC channel model can be further adapted for both monostatic and bistatic sensing scenarios. The results reveal that the communication channel and sensing channel show different propagation properties and indicates that the sensing channel is more sensitive to mobile targets. The well agreement between theoretical and simulated as well as measurement results validate the effectiveness of the proposed model.

As a promising new enable technology in 6G network, the channel modeling research on ISAC still has a long way to go. Due to the space limitation, more detailed derivations of the proposed ISAC channel model, discussions on the propagation characteristics as well as the cross-correlations between communication and sensing channels are left in the extended version of this paper.


\end{document}